\documentclass[english,pra,reprint,superscriptaddress,tightenlines,10pt]{revtex4-1}
\usepackage[T1]{fontenc}
\usepackage[latin9]{inputenc}
\usepackage{babel}
\usepackage{amsmath}
\usepackage{amssymb}
\usepackage{graphicx}
\usepackage[unicode=true]{hyperref}
\usepackage{breakurl}

\makeatletter
\hypersetup{
    colorlinks,
}
\makeatother

\begin{document}

\title{Unidirectional spin transport of a spin-orbit-coupled atomic matter wave
 using a moving Dirac $\delta$-potential well}
\author{Jieli Qin}
\email{104531@gzhu.edu.cn; qinjieli@126.com}
\address{School of Physics and Electronic Engineering, Guangzhou University,
230 Wai Huan Xi Road, Guangzhou Higher Education Mega Center, Guangzhou
510006, People\textquoteright s Republic of China}
\author{Lu Zhou}
\email{lzhou@phy.ecnu.edu.cn}
\address{Department of Physics, School of Physics and Electronic Science,
East China Normal University, Shanghai 200241, People\textquoteright s
Republic of China}

\address{Collaborative Innovation Center of Extreme Optics, Shanxi University,
Taiyuan, Shanxi 030006, People\textquoteright s Republic of China}

\begin{abstract}
We study the transport of a spin-orbit-coupled atomic matter wave using a
moving Dirac $\delta$-potential well. In a spin-orbit-coupled system, bound
states can be formed in both ground and excited energy levels with a Dirac
$\delta$-potential. Because Galilean invariance is broken in a spin-orbit-coupled
system, moving of the potential will induce a velocity-dependent
effective detuning. This induced detuning breaks the spin symmetry and makes
the ground-state transporting channel be spin-$\uparrow $ ($\downarrow $)
favored while makes the excited state transporting channel be spin-$\downarrow $
($\uparrow $) favored for a positive-direction (negative-direction) transporting. When
the $\delta$-potential well moves at a small velocity, both the ground-state
and excited-state channels contribute to the transportation, and thus both
the spin components can be efficiently transported. However, when the
moving velocity of the $\delta$-potential well exceeds a critical value, the
induced detuning is large enough to eliminate the excited bound state,
and makes the ground bound state the only transporting channel, in which only the
spin-$\uparrow $ ($\downarrow $) component can be efficiently transported in a
positive (negative) direction. This work demonstrates a prototype of
unidirectional spin transport.
\end{abstract}

\maketitle

\selectlanguage{english}

\address{School of Physics and Electronic Engineering, Guangzhou University,
230 Wai Huan Xi Road, Guangzhou Higher Education Mega Center, Guangzhou
510006, People\textquoteright s Republic of China}

\selectlanguage{english}

\address{Department of Physics, School of Physics and Electronic Science,
East China Normal University, Shanghai 200241, People\textquoteright s
Republic of China}

\address{Collaborative Innovation Center of Extreme Optics, Shanxi University,
Taiyuan, Shanxi 030006, People\textquoteright s Republic of China}

\section{Introduction\label{sec:Introduction}}

Transport of matterwave is essential in many ultracold atom physics
experiments and applications. With such a technique, an ultracold atom
experiment can be split into matterwave producing and matterwave using
modules. Each module can be optimized separately. Thus the experiment can be
performed more efficiently %
\citep{Gustavson2001Transport,Greiner2001Magnetic,Goldwin2004Measurement}.
In this thought, for example, atomic gases have been loaded into optical
cavities %
\citep{Sauer2004Cavity,Brennecke2007Cavity,Colombe2007Strong,
Culver2016Collective,Jiang2019Efficiently,Bowden2019APyramid}
and hollow fibers %
\citep{Bajcsy2009Efficient,Vorrath2010Efficient,Bajcsy2011Laser,
Hilton2018High,Yoon2019Laser}%
, leading to a good many interesting research works (for reviews see
references \citep{Ritsch2013Clod,Zhou2013Cavity,Adnan2017Experimental}). It
is also found that the transport of atomic matterwave can be very useful in the
realization of atom interferometry %
\citep{Arndt2012Focus,Eckel2014Interferometric,Xin2018AnAtom}, atomtronics
device %
\citep{Ryu2013Experimental,Ryu2015Integrated,Li2016Superfluid,Amico2017Focus}
and continuous atom laser %
\citep{Chikkatur2002Acontinuous,Lahaye2006Transport,Guerin2006Guided}.

Many schemes to transport cold atomic matterwave have been demonstrated. In
moving molasses technique \citep{Kasevich1991Atomic,Gibble1993Laser}, the
cloud of cold atoms freely flies to the destination by itself.
Controlled transport can be realized by applying an atomic waveguide %
\citep{Renn1995Laser,Ito1996Laser,Muller1999Guiding,Denschlag1999Guiding,
Key2000Propagation,Strecker2002Formation,Leanhardt2002Propagation,
Plaja2002Expansion,Gupta2005Bose,BravoAbad2006Photonic,Marchant2013Controlled}%
. However, using these techniques, the cloud of atoms expands, and its
density drops during the transporting process. To overcome this blemish,
matterwave transport using a moving potential well is introduced, and soon becomes
widely used in cold atom experiments %
\citep{Gustavson2001Transport,Greiner2001Magnetic,Goldwin2004Measurement,
Hansel2001Magnetic,Lahaye2006Transport,Schmid2006Long,Couvert2008Optimal,
Alberti2009Engineering,Roy2017AMinimalistic,Chong2018Observation}.

The moving dynamic of spin-orbit (SO) coupled matterwave shows new features.
For a SO coupled system, the Galilean invariance does not hold any more %
\citep{Zhou2012Opposite,Vyasanakere2012Collective,Zhang2016Properties}.
Different moving directions or speeds can have very different effects on the
dynamics of SO coupled atomic gases. As a result, many interesting phenomena
arise. A few examples are listed below. The critical velocity of
superfluidity becomes reference frame dependent \citep{Zhu2012Exotic}. An
oscillation of magnetization in SO coupled Bose-Einstein condensate (BEC) is
induced by the moving \citep{Zhang2012Collective,Li2012sum}. The shape of a
SO coupled BEC bright soliton changes with its velocity \citep{Xu2013Bright}%
. In a translating optical lattice, SO coupled BEC behaves anisotropically
depending on the direction of translation \citep{Hamner2015Spin}. The normal
density of superfluidity does not vanish even at zero temperature %
\citep{Zhang2016Superfluid}. And, non-magnetic one-way spin switch %
\citep{Mossman2019Experimental} and spin-current generation %
\citep{Li2019Spin} have also been demonstrated recently.

In this paper, we study the transport of SO coupled cold atomic matterwave
using a moving Dirac delta-potential well, see diagram figure
\ref{fig:Diagram}. We show that the delta-potential well can at most support
both a ground and an excited bound state in SO coupled cold atom system.
These bound states can be used to efficiently transport the cold atomic
matterwave, thus serve as the transporting channels in the problem. Moving
of the potential well will induce a velocity proportional effective
detuning, which can substantially affect the transporting channels. This
induced detuning breaks the spin symmetry, and makes the two transporting
channels being spin-polarized. For a slowly positive direction moving
delta-potential well, the ground state channel is spin-$\uparrow$ favored,
while the excited state channel is spin-$\downarrow$ favored. Both the spin-$%
\uparrow$ and spin-$\downarrow$ components can be transported through its
favorable channel. And for a slowly negative direction moving
delta-potential well, things are very similar, except that the roles of spin-%
$\uparrow$ and spin-$\downarrow$ exchange with each other. When the velocity
of the moving delta-potential well exceeds a critical value, the induced
effective detuning will be large enough to lift the excited state out of the
binding ability of the delta-potential well, thus eliminate the excited bound
state transporting channel. Therefore, in such a case,
matterwave can only be transported
through the ground state channel. Since for a positive (negative) direction
moving delta-potential well, the ground state channel is spin-$\uparrow$
(spin-$\downarrow$) dominant, only this appropriate spin component of matterwave
can be efficiently transported. These unidirectional transporting features
indicate that the system considered here may potentially be used to realize
spintronic devices such as spin diode \citep{Cheuk2012SpinInjection,Lan2015SpinWave},
valve \citep{Zhao2019Tunalble} and filter \citep{Lebrat2019Quantized,Corman2019Quantized}.

The rest part of this paper is organized as follows: In section \ref%
{sec:Model}, the physical model of this paper is presented. In section \ref%
{sec:TransportingChannels}, we solve the bound states (i.e., the
transporting channels) of the moving delta-potential well, and discuss their
properties. In section \ref{sec:Results}, the transporting dynamics and
efficiency are shown. And at last, the paper is summarized in section \ref%
{sec:Summary}.

\section{Model\label{sec:Model}}

\begin{figure}[tbp]
\begin{centering}
\includegraphics{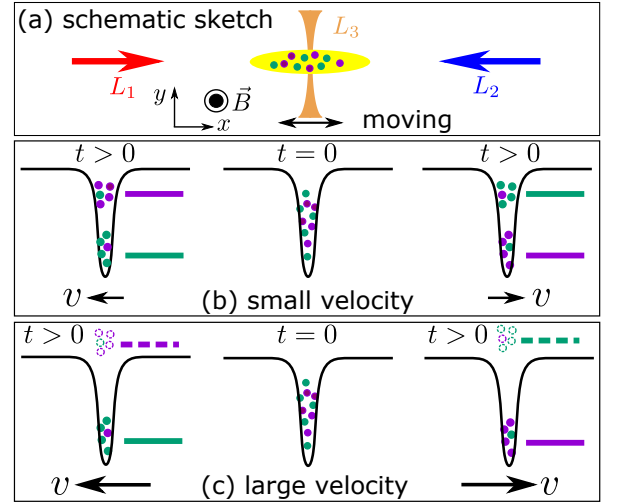}
\par\end{centering}
\caption{Diagram of transporting SO coupled cold atomic matterwave using a
moving delta-potential well. At time $t=0$, an atomic SO coupled BEC is
prepared in the ground state of a delta-potential well. The atoms are
equally distributed in the spin-$\uparrow$ and spin-$\downarrow$ components.
Afterward, one moves the delta-potential well at velocity $v$ to transport
the matterwave.
Top panel (a): Schematic sketch of the system. The SO coupling
is realized by the two counter-propagating Raman lasers $L_1$ and $L_2$. The
third tightly focused movable laser beam $L_3$ generates the delta-potential well.
Middle panel (b): For a small velocity moving delta-potential
well, it can support two transporting channels --- ground and excited bound
states of the moving delta-potential well in the comoving frame. One of the
transporting channels is spin-$\uparrow$ favored, while the other one is
spin-$\downarrow$ favored. Both the spin-$\uparrow$ and spin-$\downarrow$
components can be efficiently transported in such a case. Bottom panel (c):
For a large velocity moving delta-potential well, the moving induced
effective detuning lifts the excited state out of the binding ability of the
potential well, thus matterwave transportation can only take place in the
ground state channel. For a positive (negative) direction moving
delta-potential well, the ground state channel is spin-$\uparrow$ (spin-$%
\downarrow$) dominant, thus only the spin-$\uparrow$ (spin-$\downarrow$)
component can be efficiently transported. }
\label{fig:Diagram}
\end{figure}

We consider the transporting of quasi-one-dimensional SO coupled cold atoms
using a moving delta-potential well, see figure \ref{fig:Diagram}. The SO
coupling is realized by the two x-direction counter-propagating Raman lasers
$L_1$ and $L_2$ \citep{Lin2011Spin}. And the delta-potential well can be
generated using a y-direction shining tightly focused laser beam
\citep{Uncu2007Bose,Garrett2011Growth}.
Such a system can be described by Hamiltonian
\begin{align}
H & =H_{0}+U\left(x,t\right),  \label{eq:Hamiltonian}
\end{align}
where $U\left(x,t\right)$ is the external potential, and $H_{0}$ is the SO
coupled free particle Hamiltonian
\begin{equation}
H_{0}=\left[%
\begin{matrix}
\frac{\left(p_{x}-p_{c}\right)^{2}}{2m}-\frac{\hbar\Delta_{0}}{2} & \frac{%
\hbar\Omega}{2} \\
\frac{\hbar\Omega}{2} & \frac{\left(p_{x}+p_{c}\right)^{2}}{2m}+\frac{%
\hbar\Delta_{0}}{2}%
\end{matrix}%
\right].  \label{eq:Hamiltonian0}
\end{equation}
Here $p_{x}=\hbar k_{x}=-i\hbar\frac{\partial}{\partial x}$ is the
one-dimensional momentum operator, $p_{c}=\hbar k_{c}$ is the strength of SO
coupling determined by momentum transfer during the Raman scattering
process, $\Delta_{0}$ is the detuning of the Raman driving from the atomic
energy level splitting (in this paper we assume that its value is set to zero
$\Delta_{0}=0$), $\Omega$ is the effective Rabi frequency for Raman flipping
between the two spin states. The inter-atom collision interaction is not
included here, as it is assumed to have been eliminated by the Feshbach
resonance technique \citep{Chin2010Feshbach,Timmermans1999Feshbach}. In the
following contents, for convenience natural unit $\hbar=m=1$ will be used.

Before time $t=0$, the system is prepared in the ground state of a
delta-potential well localized at $x=0$. Then, for time $t>0$, we move the
delta-potential well at a constant velocity $v$, and study the subsequent
transportation. Hence, the external potential $U\left(x,t\right)$ can be
written in a piece-wise function as follows
\begin{equation}
U\left(x,t\right)=%
\begin{cases}
V_{0}\left(x\right)=-V_{0}\delta\left(x\right), & t\leq0, \\
V\left(x,t\right)=-V_{0}\delta\left(x-vt\right), & t>0.%
\end{cases}
\label{eq:potential}
\end{equation}
The initial state can be constructed using the free particle oscillating
evanescent wave modes (for detail see section \ref{sec:TransportingChannels}%
)
\begin{equation}
\psi_{0}=%
\begin{cases}
A_{0,1}\xi_{-;1}e^{ik_{x;1}x}+A_{0,2}\xi_{-;2}e^{ik_{x;2}x}, & x\leq0, \\
A_{0,3}\xi_{-;3}e^{ik_{x;3}x}+A_{0,4}\xi_{-;4}e^{ik_{x;4}x}, & x>0.%
\end{cases}
\label{eq:psi0}
\end{equation}
The transporting dynamic for time $t>0$ is governed by time-dependent Schr\"{o}dinger equation
\begin{equation}
i\frac{\partial\psi\left(x,t\right)}{\partial t}=H\left(x,t\right)\psi%
\left(x,t\right).  \label{eq:Schrodinger}
\end{equation}
Noticed that here the Hamiltonian is time-dependent, it will be convenient
to deal with the problem in a frame comoving with the potential. So we take
the following transformation
\begin{equation}
x\rightarrow x-vt,\quad t\rightarrow t,  \label{eq:transformation_12}
\end{equation}
and
\begin{equation}
\psi\rightarrow\psi e^{-ivx}e^{iv^{2}t/2}.  \label{eq:transformation_3}
\end{equation}
Under this transformation, equation (\ref{eq:Schrodinger}) becomes
\begin{equation}
i\frac{\partial\psi\left(x,t\right)}{\partial t}=H_{t}\psi\left(x,t\right),
\label{eq:MovintgSchordinger}
\end{equation}
with the transformed Hamiltonian being \citep{Zhang2012Collective}

\begin{align}
H_{t} & =\frac{1}{2}\left[%
\begin{matrix}
\left(k_{x}-k_{c}\right)^{2}-\Delta & \Omega \\
\Omega & \left(k_{x}+k_{c}\right)^{2}+\Delta%
\end{matrix}%
\right]+V_{0}\left(x\right),  \label{eq:Hamiltonian_t}
\end{align}
where
\begin{equation}
\Delta=2k_{c}v, \label{eq:effectiveDetuning}
\end{equation}
is an effective detuning induced by the moving of
external potential.

This new time-independent Hamiltonian is similar to the Hamiltonian at time $%
t=0$, except for the moving induced additional detuning term. Therefore, the
moving delta-potential also supports bound states, and these bound states
can also be constructed similarly using the oscillating evanescent waves
(for detail also see section \ref{sec:TransportingChannels})
\begin{equation}
\psi_{tb}=%
\begin{cases}
A_{t,1}\xi_{-;t1}e^{ik_{x;t1}x}+A_{t,2}\xi_{-;t2}e^{ik_{x;t2}x}, & x\leq0,
\\
A_{t,3}\xi_{-;t3}e^{ik_{x;t3}x}+A_{t,4}\xi_{-;t4}e^{ik_{x;t4}x}, & x>0.%
\end{cases}
\label{eq:psitb}
\end{equation}
These states are bounded by, and at the same time, comove with the external
potential, thus can serve as the transporting channels of the system. In the
next section, we will show that the moving delta-potential well can at most
support two bound states. We label them as $\psi_{tb;g}$ and $\psi_{tb;e}$,
with subscript ``$g$'' and ``$e$'' meaning the ground and excited states.
And, the excited state disappears for large potential moving velocity.
Hamiltonian (\ref{eq:Hamiltonian_t}) also supports an infinite number of
scattering states. However, after a long-time evolution, these states spread
all over the whole space and have negligible densities. Therefore, they are
not important for the transportation. Neglecting them, the efficiently
transported matterwave can be described by the following wavefunction %
\citep{Granot2009Quantum,Sonkin2010Trapping}

\begin{equation}
\psi_{t}=C_{g}\psi_{tb;g}e^{-iE_{g}t}+C_{e}\psi_{tb;e}e^{-iE_{e}t},
\label{eq:psit}
\end{equation}
where $E_{g}$ and $E_{e}$ are the ground state and excited state
eigenenergies, and $C_{g}$ and $C_{e}$ are the ground state and excited
state probability amplitudes determined by the initial wavefunction
according to formulae
\begin{equation}
C_{g,e}=\int_{-\infty}^{\infty}\psi_{tb;g,e}^{\dagger}\psi_{0}e^{ivx}dx.
\label{eq:Cge}
\end{equation}
Here, when the excited state does not exist, one simply sets $C_{e}=0$ to
eliminate its role.

At last, we define some quantities to characterize the transporting
efficiency of the moving delta-potential well. The time averaged spin-$%
\uparrow$ and spin-$\downarrow$ atom numbers of transported matterwave are
given by
\begin{equation}
N_{t;\uparrow,\downarrow}=\sum_{i=g,e}\int\left|C_{i}\psi_{tb;i\uparrow,%
\downarrow}\right|^{2}dx,  \label{eq:NtUpDown}
\end{equation}
And the time averaged total atom number of transported matterwave is
\begin{equation}
N_{t}=N_{t;\uparrow}+N_{t;\downarrow}=\left|C_{g}\right|^{2}+\left|C_{e}%
\right|^{2}.  \label{eq:Nt}
\end{equation}
We emphasize that the total atom number of initial state $\psi_{0}$ will be
normalized to $1$ in this paper, therefore $N_{t;\uparrow\downarrow}$ and $%
N_{t}$ defined here indeed can be interpreted as the fraction of transported atoms
compared to the initial state.

\section{Transporting channels: bound states of the moving delta-potential
well\label{sec:TransportingChannels}}

\begin{figure*}[tbp]
\begin{centering}
\includegraphics{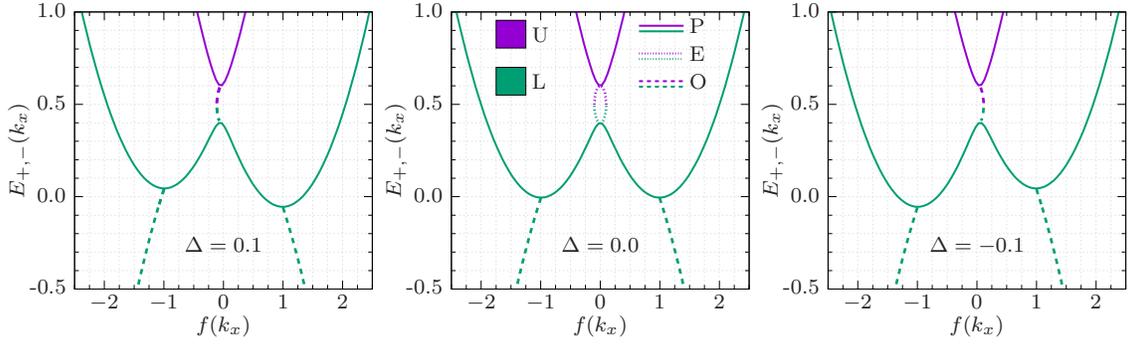}
\par\end{centering}
\caption{Free particle spectrum of SO coupled cold atomic matterwave. SO
coupling and Rabi coupling strengths are $k_{c}=1,\Omega=0.2$. Left panel:
positive detuning $\Delta=0.1$. Middle panel: zero detuning $\Delta=0$.
Right panel: negative detuning $\Delta=-0.1$. The violet lines correspond to
the upper (``U'') spectrum branch $E_{+}$, while the green lines correspond
to the lower (``L'') spectrum branch $E_{-}$. The solid, dashed and dotted
lines represent plane traveling (``P''), oscillating evanescent (``O'') and
ordinary evanescent (``E'') waves, respectively. For plane traveling wave,
wavevector $k_{x}$ has a real number value, the $x$-axis is set to $%
f\left(k_{x}\right)=k_{x}$; for oscillating evanescent wave, $k_{x}=\protect%
\beta\pm i\protect\alpha$, the $x$-axis is set to $f\left(k_{x}\right)=%
\mathrm{sgn}\left[\mathrm{Re}\left(k_{x}\right)\right]\cdot\left|k_{x}\right|
$ (therefore, the lines for $+\protect\alpha$ and $-\protect\alpha$ overlap
with each other); and for ordinary evanescent wave, $k_{x}=\pm i\protect%
\alpha$, the $x$-axis is set to $f\left(k_{x}\right)=\mathrm{sgn}\left[%
\mathrm{Im}\left(k_{x}\right)\right]\cdot\left|k_{x}\right|$. }
\label{fig:FreeParticleSpectrum}
\end{figure*}

From the previous section, one sees that both the initial state and
transporting channel problems involve finding the bound eigenstates of SO
coupled cold atomic matterwave. For a delta-potential well $%
V_{0}\left(x\right)=-V_{0}\delta\left(x\right)$, it is equivalent to a free
space problem except for the very point $x=0$. So the bound eigenstates of a
delta-potential well can be constructed using the free particle modes by
matching boundary conditions at $x=0$ \cite{Qin2020Bound}. In this section,
we first discuss the free particle spectrum and eigenstates of a SO coupled
system, and then construct the bound states using the free particle modes.

The free particle modes can be found by diagonalizing Hamiltonian
(\ref{eq:Hamiltonian_t}) (with $V_0\left(x\right)$ neglected).
It comes out that the eigenenergy is
given by
\begin{equation}
\left[E-\frac{\left(k_{x}^{2}+k_{c}^{2}\right)}{2}\right]^{2}-%
\left(k_{c}k_{x}+\frac{\Delta}{2}\right)^{2}+\left(\frac{\Omega}{2}%
\right)^{2}=0,  \label{eq:eigenvalues}
\end{equation}
which is a second-order equation of $E$. This indicates that the spectrum
will split into two branches (we recognize them as ``lower'' and ``upper''
branch in this paper)
\begin{equation}
E_{\pm}=\frac{k_{x}^{2}+k_{c}^{2}}{2}\pm\frac{1}{2}\sqrt{\left(2k_{c}k_{x}+%
\Delta\right)^{2}+\Omega^{2}},  \label{eq:spectrum}
\end{equation}
and the corresponding eigenstates are
\begin{equation}
\psi_{\pm}\left(k_{x}\right)=\xi_{\pm}\left(k_{x}\right)e^{ik_{x}x}=C_{\pm}%
\left(%
\begin{matrix}
\zeta_{\pm} \\
1%
\end{matrix}%
\right)e^{ik_{x}x},  \label{eq:eigenstates}
\end{equation}
where $\zeta_{\pm}=-\left(2k_{c}k_{x}+\Delta\right)/\Omega\pm\sqrt{%
\left(2k_{c}k_{x}+\Delta\right)^{2}/\Omega^{2}+1}$ characterize the spin
wavefunction, and $C_{\pm}=1/\sqrt{1+\left|\zeta_{\pm}\right|^{2}}$ is the
normalization parameter.

Equation (\ref{eq:eigenvalues}) admits three kinds of wavevectors with real,
pure imaginary and complex value, respectively. The real part of wavevector $%
k_{x}$ contributes a plane traveling wave factor in the eigenstate (\ref%
{eq:eigenstates}), while the imaginary value part contributes an exponential
decay factor. Thus, corresponding to these three kinds of wavevectors, the
eigenstates are plane traveling wave, ordinary evanescent wave, and
oscillating evanescent wave states, respectively %
\citep{Sablikov2007Evanescent,Zhou2015Goos}. For different strengths of SO
coupling $k_{c}$ and Rabi coupling $\Omega$, the spectrum of the system can
be a little different. Here we focus on the strong SO coupling case with $%
k_{c}>\Omega/2$, and numerically we choose $k_{c}=1$ and $\Omega=0.2$ all
over the paper. In figure \ref{fig:FreeParticleSpectrum}, the free particle
spectra are plotted for different detunings $\Delta=0,\pm0.1$. In the
figure, different types of eigenstates are marked with different styles of
lines. From the figure, one sees that for energy below the minimum of lower
branch spectrum $E_{0}<E_{-,min}$, the four eigenmodes are all oscillating
evanescent waves. Further calculations show that the corresponding
wavevectors, i.e., solutions of equation (\ref{eq:eigenvalues}), have the
following symmetric form
\begin{equation}
k_{x;1,3}=\beta\mp i\alpha_{\text{I}},\quad k_{x;2,4}=-\beta\mp i\alpha_{%
\text{II}}.  \label{eq:kx12kx34}
\end{equation}
Here $\alpha_{\text{I,II}}$ and $\beta$ are real positive numbers.
Specially, when $\Delta=0$, the two imaginary part numbers also equal each
other $\alpha_{\text{I}}=\alpha_{\text{II}}=\alpha$.

Since waves $\exp\left[ik_{x;1,2}\right]=\exp\left[\pm i\beta x+\alpha_{%
\text{I,II}}x\right]$ decay to zero when $x\rightarrow-\infty$, while waves $%
\exp\left[ik_{x;3,4}\right]=\exp\left[\pm i\beta x-\alpha_{\text{I,II}}x%
\right]$ decay to zero when $x\rightarrow+\infty$, the bound states of a
delta-potential well can be constructed using these four oscillating
evanescent wave modes. The wave function can be written as
\begin{align}
\psi_{b} & =%
\begin{cases}
A_{1}\xi_{-;1}e^{ik_{x;1}x}+A_{2}\xi_{-;2}e^{ik_{x;2}x}, & x<0, \\
A_{3}\xi_{-;3}e^{ik_{x;3}x}+A_{4}\xi_{-;4}e^{ik_{x;4}x}, & x>0,%
\end{cases}
\notag \\
& =%
\begin{cases}
A_{1}\xi_{-;1}e^{i\beta x+\alpha_{\text{I}}x}+A_{2}\xi_{-;2}e^{-i\beta
x+\alpha_{\text{II}}x}, & x<0, \\
A_{3}\xi_{-;3}e^{i\beta x-\alpha_{\text{I}}x}+A_{4}\xi_{-;4}e^{-i\beta
x-\alpha_{\text{II}}x}, & x>0,%
\end{cases}
\label{eq:psib}
\end{align}
with symbols $\xi_{-;1,2,3,4}=\xi_{-}\left(k_{x;1,2,3,4}\right)$ for
shorthand. The wavefunction parameters $A_{1,2,3,4}$ and eigenenergy $E_{b}$
($k_{x;1,2,3,4}$ are determined by $E_{b}$ according to equation (\ref%
{eq:eigenvalues})) are to be determined by normalization constrain $%
\int_{-\infty}^{\infty}\left|\psi_{b}\right|^{2}dx=1$ together with boundary
conditions: continuity of wavefunction
\begin{equation}
\left.\psi_{b}\left(x\right)\right|_{0+}=\left.\psi_{b}\left(x\right)%
\right|_{0-},  \label{eq:boudnary1}
\end{equation}
and jump of the first-order derivative of wave function caused by the
singularity of delta-potential
\begin{equation}
\left.\frac{d\psi_{b}\left(x\right)}{dx}\right|_{0+}-\left.\frac{%
d\psi_{b}\left(x\right)}{dx}\right|_{0-}=-2V_{0}\psi_{b}\left(x=0\right).
\label{eq:boudnary2}
\end{equation}

\begin{figure}[tbp]
\begin{centering}
\includegraphics{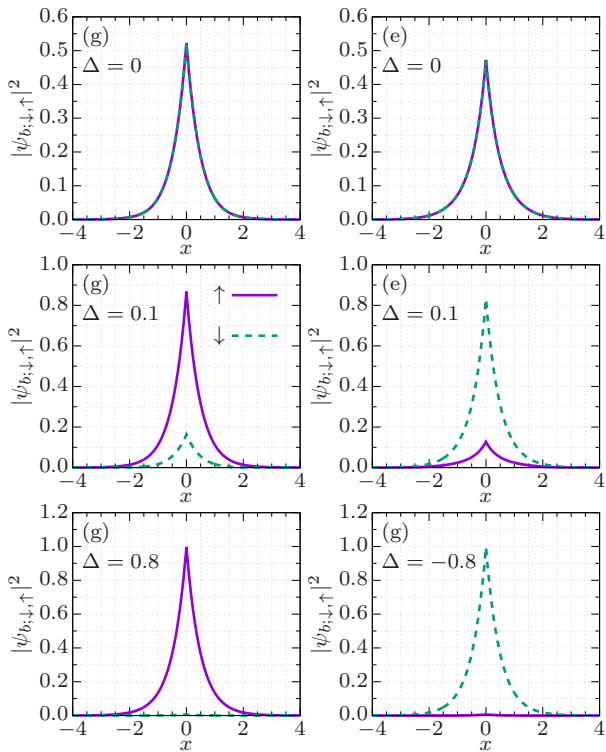}
\par\end{centering}
\caption{Eigenstates of delta-potential well trapped SO coupled cold atomic
matterwave for different effective detunings ($\Delta=2k_{c}v=0,0.1,\pm 0.8$). The solid
violet lines stand for spin-$\uparrow$ component atomic density $\left|%
\protect\psi_{b;\uparrow}\right|^{2}$, while the green dashed lines stand
for spin-$\downarrow$ component atomic density $\left|\protect\psi%
_{b;\uparrow}\right|^{2}$. Top panels: spin symmetric ground (``g'') and
excited (``e'') states for zero detuning $\Delta=0$. Middle panels: spin
asymmetric ground and excited states for a small detuning $\Delta=0.1$.
Bottom panels: spin asymmetric ground states for large positive $\Delta=0.8$
and negative detuning $\Delta=-0.8$. The excited state is absent for such
large detunings, thus only the ground state is plotted. The eigenenergies
corresponding to these states are $-0.5536, -0.4540, -0.5742, -0.4334$,
-0.9056, -0.9056, respectively. The SO coupling and Rabi coupling strengths
are $k_{c}=1,\Omega=0.2$, and the depth of the delta-potential well is $%
V_{0}=1$. }
\label{fig:Eigenstates}
\end{figure}
\begin{figure}[tbp]
\begin{centering}
\includegraphics{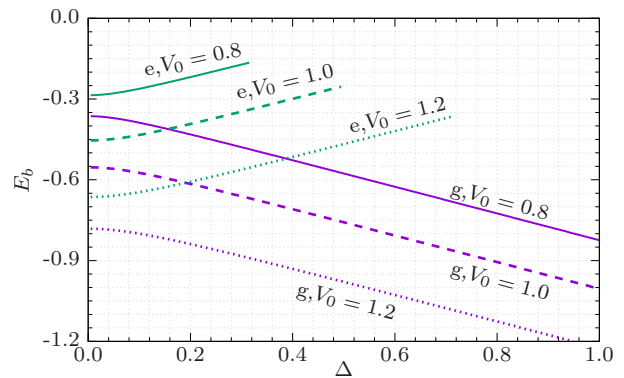}
\par\end{centering}
\caption{The spectrum of delta-potential bounded SO coupled cold atomic
matterwave. Ground (``g'', violet lines) and excited (``e'', green lines)
state energies $E_{b;g,e}$
are plotted as a function of effective detuning $\Delta=2k_{c}v$. Solid,
dashed and dotted lines represent different potential well depths $%
V_{0}=0.8,1.0,1.2$, respectively. SO coupling strength and Rabi frequency
are $k_{c}=1,\Omega=0.2$. }
\label{fig:BoundedSpectrum}
\end{figure}

\begin{figure}[tbp]
\begin{centering}
\includegraphics{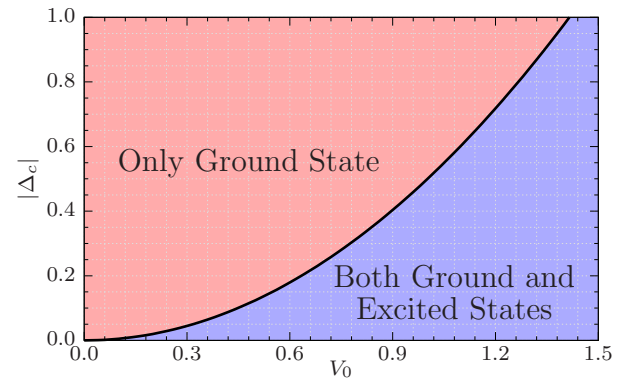}
\par\end{centering}
\caption{Critical detuning for the disappearing of the excited state. The
absolute value of critical detuning is plotted as a function of
delta-potential well depth $V_{0}$, the black solid line. Below this line,
the light blue color filled region supports both a ground and an excited
bound state. While above this line, the pink color filled region supports
only one bound state, the ground state. SO coupling strength and Rabi
frequency are $k_{c}=1,\Omega=0.2$. }
\label{fig:CriticalDetuning}
\end{figure}
\begin{figure*}[tbp]
\begin{centering}
\includegraphics{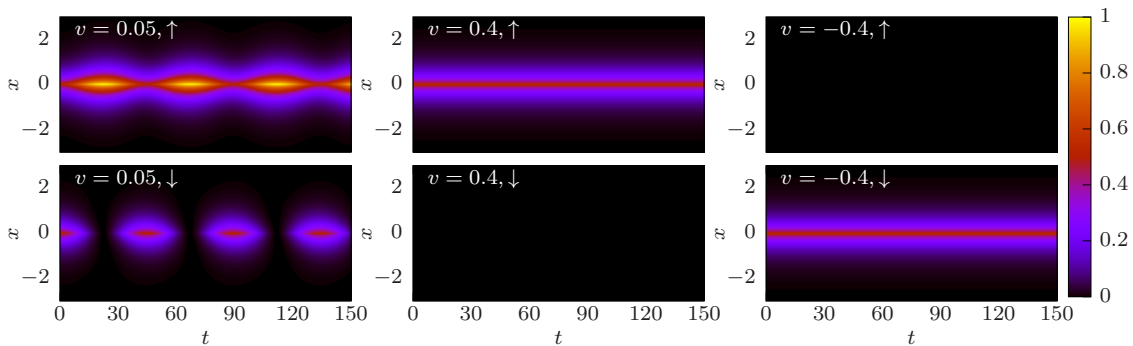}
\par\end{centering}
\caption{Time evolution of the transported matterwave under different
potential moving velocities. Left panel: Interference pattern between the
ground state and excited state channels transported matterwave for a small
velocity ($v=0.05$) moving delta-potential well. Middle penal: Spin-$\uparrow
$ component dominant transporting for a large velocity positive direction
moving ($v=0.4$) delta-potential well. Right panel: Spin-$\downarrow$
component dominant transporting for a large velocity negative direction
moving ($v=-0.4$) delta-potential well. SO coupling strength, Rabi frequency
and depth of the delta-potential well are $k_{c}=1,\Omega=0.2$ and $V_{0}=1.0
$ for all plots. }
\label{fig:TimeEvolution}
\end{figure*}
\begin{figure}[tbp]
\begin{centering}
\includegraphics{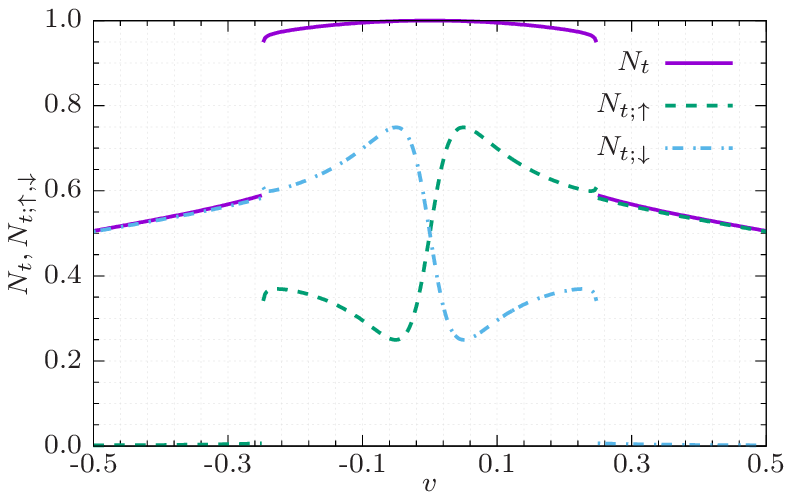}
\par\end{centering}
\caption{Amount of transported atomic matterwave for different transporting
velocities. The violet solid line represents the total amount of transported
matterwave $N_{t}$ defined in equation (\protect\ref{eq:Nt}). The green
dashed and cyan dash-dotted lines represent the amount of spin-$\uparrow$
and spin-$\downarrow$ components, $N_{t;\uparrow}$ and $N_{t;\downarrow}$
defined in equation (\protect\ref{eq:NtUpDown}). An abrupt dropping of the
amount of transported matterwave happens around $v\approx\pm0.25$ which is the
critical velocity for the disappearing of the excited state transporting
channel. The parameters used are $k_{c}=1.0$, $\Omega=0.2$, $V_{0}=1.0$. }
\label{fig:TransportCurve}
\end{figure}

Because of the spin-$1/2$ nature of the system, a delta-potential well can
support two bound states (which are recognized as ground and excited states
in this paper) with spin symmetry $\left|\psi_{\downarrow}\right|^{2}=\left|%
\psi_{\uparrow}\right|^{2}$ when the detuning is absent ($\Delta=0$). When a
small detuning is introduced, this spin symmetry is broken, $%
\left|\psi_{\downarrow}\right|^{2}\neq\left|\psi_{\uparrow}\right|^{2}$.
According to the Hamiltonian (\ref{eq:Hamiltonian}), a positive detuning
will raise the energy of spin-$\downarrow$ (or in other words, negative
momentum) component, and at the same time, lower the energy of spin-$\uparrow
$ (negative momentum) component. This fact can also be seen from the
spectrum shown in figure \ref{fig:FreeParticleSpectrum}. As a result, the
ground state becomes spin-$\uparrow$ favored, while the excited state
becomes spin-$\downarrow$ favored. When the detuning becomes large, the
energy of the excited state will be raised out of the binding ability of the
potential well, hence the excited state disappears. For a negative detuning,
the roles of spin-$\uparrow$ and spin-$\downarrow$ exchange with each other.
These facts are shown in figure \ref{fig:Eigenstates}, where ground and
excited bound states are plotted for zero and small detunings $\Delta=0,0.1$%
. And for large detunings $\Delta=\pm0.8$, the excited state disappears,
only the ground state is plotted.

We also studied the delta-potential well bounded SO coupled spectrum. The
ground state and excited state energies are plotted as a function of
detuning $\Delta$ in figure \ref{fig:BoundedSpectrum}. The detuning induced
energy splitting of the two states, and the disappearing of the excited
state can be clearly seen in this figure. And one also noticed that for a
deeper potential well the excited state disappears at a larger value of
detuning $\Delta$. This is further shown in figure \ref{fig:CriticalDetuning}%
, where the excited state disappearing critical detuning $\Delta_{c}$ is
plotted as a function of delta-potential well depth $V_{0}$. In this figure,
in the region below the critical-value-line (the black solid line), both the
ground and excited states exist, while above the line the excited state
disappears, and there is only one bound state, the ground state.

At last, recalling that moving can induce an effective detuning $%
\Delta=2k_{c}v$, it is concluded that a small positive (negative) velocity
moving delta-potential well can support both a spin-$\uparrow$ (spin-$%
\downarrow$) favored ground state and a spin-$\downarrow$ (spin-$\uparrow$)
favored excited state transporting channel, while a large positive
(negative) velocity moving delta-potential well can only support a spin-$%
\uparrow$ (spin-$\downarrow$) dominant ground state transporting channel.
This will lead to very different transporting properties of the
delta-potential wells moving with different velocities.

\section{Transportation\label{sec:Results}}

When the delta-potential well moves at a small velocity (moving induced
detuning fulfills $\left|\Delta\right|=2k_{c}\left|v\right|<\left|\Delta_{c}%
\right|$, with critical detuning $\left|\Delta_{c}\right|$
having been shown in figure \ref{fig:CriticalDetuning}),
both the ground state and excited state channels
participate in the transporting, and interference will happen between them.
As a result, an oscillation of the atomic density can be observed during the
transporting process. The oscillating period is determined by the energy
difference between the ground and excited states, $T=2\pi/\left(E_{e}-E_{g}%
\right)$. In the left panel of figure \ref{fig:TimeEvolution}, such an
oscillation is shown for a delta-potential well with depth $V_{0}=1.0$ moving
at velocity $v=0.05$. The ground state and excited state energies are
-0.5742, -0.4334, respectively. Therefore, the oscillating period is $%
T\thickapprox44.63$ in the figure.

When the delta-potential well moves at a large velocity ($%
\left|\Delta\right|=2k_{c}\left|v\right|>\left|\Delta_{c}\right|$), the
moving induced detuning eliminates the excited state transporting channel,
the ground state channel plays the only transporting role. Without the
excited state channel to interfere with, no oscillation happens in this
case. For a positive direction moving potential, the ground state channel is
spin-$\uparrow$ dominant, thus only the spin-$\uparrow$ component matterwave
can be efficiently transported. While for a negative direction moving
potential, the spin-$\downarrow$ dominant ground state channel can only
efficiently transport the spin-$\downarrow$ component matterwave. This is
shown in the middle and right panels in figure \ref{fig:TimeEvolution}.

We also examined the relationship between transporting efficiency
and moving velocity of the delta-potential well. The total
transported atom number $N_{t}$, and atom number of each spin component $%
N_{t;\uparrow,\downarrow}$ are plotted as a function of the potential moving
velocity $v$ in figure \ref{fig:TransportCurve}. In the figure, below the
critical velocity ($\left| v \right| < 0.25$), for a positive direction
moving potential, the spin-$\uparrow$ component is a little favored; while
for a negative direction moving potential, it is the spin-$\downarrow$
component favored. And the total transported atom number takes a value of
nearly 1 (not less than 95\%), which means that almost all the atoms can be
transported in this case. However,when the velocity of the potential well
exceeds the critical value ($\left| v \right| > 0.25$), one of the spin
components (spin-$\downarrow$ for positive, while spin-$\uparrow$ for
negative direction transporting) suddenly drops to almost 0, and only the
other spin component can still be transported. Except for this sudden
dropping of transporting efficiency, in the figure one also notices a slowly
dropping of the total transported atom number as the transporting velocity
increases. Mathematically, this is because, for a large value of $v$, the
factor $\exp\left[ivx\right]$ in equation (\ref{eq:Cge}) contributes a fast
oscillation, which will reduce the transported matterwave amplitude.
Physically, this can be explained by the fact that a faster moving of the
potential tends to excite more atoms out of the trapping well.

At last, we also point out that here the moving of the delta-potential is
switched on abruptly. However, one can also discuss the case of adiabatically
switching on. In such a case, according to the adiabatical theorem \citep{Kato1950},
the atoms will adiabatically follow the ground state of the moving potential (we
assume the atoms are initially prepared in the ground state). And as we have already
demonstrated the dependence of ground state spin polarization on potential moving
velocity in section \ref{sec:TransportingChannels}, therefore the unidirectional spin
transport can be achieved as well.

\section{Summary\label{sec:Summary}}

In summary, we have studied the transport of SO coupled cold atomic matterwave
using a moving Dirac delta-potential well. The transporting can happen in
two different channels (the ground and excited bound states of the moving
delta-potential well). Due to that SO coupling breaks Galilean invariance,
the transport shows a prominent unidirectional property. For small
moving velocity, both the ground state and excited state channels contribute
to the transportation, and the two spin components can both be efficiently
transported, where spin-$\uparrow $ (spin-$\downarrow $) is a little favored
for a positive (negative) direction transport. And under such a case, the
interference between the ground state and excited state channels will cause
an oscillation of the transported matterwave density. When the moving
velocity exceeds a critical value, the excited state transporting channel
disappears, only one spin component of the matterwave can be efficiently
transported through the ground state channel. Positive direction moving
delta-potential well only efficiently transports spin-$\uparrow $ component,
while negative direction moving potential well only efficiently transports
spin-$\downarrow $ component. The critical moving velocity is also
identified in the paper. Note that some experimentally realizable potentials %
\citep{Uncu2007Bose,Garrett2011Growth,Lacki2016Nanoscale,Wang2018Dark} can
be modeled by delta-function, the phenomena reported here are expected to be
observed experimentally.
\begin{acknowledgments}
This work is supported by National Natural Science Foundation of China
(Grant Nos. 11904063, 11847059 and 11374003).
\end{acknowledgments}

\end{document}